\documentclass{article}

\usepackage{arxiv}
\usepackage[utf8]{inputenc} 
\usepackage[T1]{fontenc}    
\usepackage{hyperref}       
\usepackage{url}            
\usepackage{booktabs}       
\usepackage{amsfonts}       
\usepackage{nicefrac}       
\usepackage{microtype}      
\usepackage{lipsum}		    
\usepackage{graphicx}
\usepackage[dvipsnames]{xcolor} 
\usepackage{natbib}
\usepackage{doi}
\usepackage{multirow}
\usepackage{appendix}
\usepackage{array}              
\usepackage{flafter}            
\usepackage{float}
\usepackage{caption}            
\usepackage{subcaption}         
\usepackage{amsmath}            

\DeclareCaptionLabelFormat{boldfig}{\textbf{Fig.\ #2}}
\hypersetup{
    colorlinks=true,
    linkcolor=Black,
    filecolor=RoyalPurple,
    urlcolor=RoyalPurple,
    citecolor=RoyalPurple,
}

\title{Synthetic Protein-Ligand Complex Generation for Deep Molecular Docking}
\author{
    Sofiene Khiari
    \thanks{Department of Pharmaceutical Sciences, University of Basel, Basel, Switzerland}
    \ \thanks{Swiss Institute of Bioinformatics, Basel, Switzerland}
    \ \thanks{Equal contribution}
    \And
    Matthew R. Masters \footnotemark[1] \ \footnotemark[2] \ \footnotemark[3] \And
    Amr H. Mahmoud \footnotemark[1] \ \footnotemark[2] \And
    Markus A. Lill \footnotemark[1] \ \footnotemark[2] \And
}
\date{}

\begin{document}
\maketitle


\begin{abstract}

The scarcity of experimental protein-ligand complexes poses a significant challenge for training robust deep learning models for molecular docking.
Given the prohibitive cost and time constraints associated with experimental structure determination, scalable generation of realistic protein-ligand complexes is needed to expand available datasets for model development.
In this study, we introduce a novel workflow for the procedural generation and validation of synthetic protein-ligand complexes, combining a diverse ensemble of generation techniques and rigorous quality control.
We assessed the utility of these synthetic datasets by retraining established docking models, Smina and Gnina, and evaluating their performance on standard benchmarks including the PDBBind core set and the PoseBusters dataset.
Our results demonstrate that models trained on synthetic data achieve performance comparable to models trained on experimental data, indicating that current synthetic complexes can effectively capture many salient features of protein-ligand interactions.
However, we did not observe significant improvements in docking or scoring accuracy over conventional methods or experimental data augmentation.
These findings highlight the promise as well as the current limitations of synthetic data for deep learning-based molecular docking and underscore the need for further refinement in generation methodologies and evaluation strategies to fully exploit the potential of synthetic datasets for this application.

\end{abstract}

\section{Introduction}
\label{introduction}

Protein-ligand docking is an important task in drug discovery and design that aims to predict the binding mode of a ligand to its protein receptor.
Docking is widely-used and can be utilized in many different workflows including rational drug design, virtual screening, and toxicity prediction \cite{sethi2019molecular}.
Traditionally, docking has been done using classical sampling algorithms such as Monte Carlo or evolutionary algorithms paired with a handcrafted scoring function based on physical principles or statistical models \cite{Halperin2002PrinciplesOD}.
While this approach is powerful and widely employed, it also has inherent limitations.
For example, the sampling process can be computationally expensive, especially for large and flexible systems, and can have difficulties overcoming potential energy barriers to effectively sample the correct pose.
Scoring functions also have limited accuracy and can often misidentify the preferred binding mode \cite{Clark2016PredictionOP}.

In hopes of addressing these shortcomings, a new generation of docking methods utilizing deep learning (DL) models have started to replace the search and scoring functions \cite{Vittorio2024AddressingDP}.
However, the amount of protein-ligand training data is severely limited and unbalanced, leading to models that overfit and are unable to generalize to unseen new molecules \cite{Ellingson2020MachineLA}.
This problem will continue to persist in the future, due to the large cost and time investment associated with experimental structure determination.
Additionally, most of the protein-ligand complex space is unknown and has not yet been explored, or simply cannot be explored due to technical limitations \cite{COLWELL2018123}.
It has been seen in other areas of deep learning, such as image generation and large-language models, that there is an immense benefit to learning from a plethora of available data \cite{Schuhmann2022LAION-5B, Li2021Grounded}.
Therefore, a method to generate realistic protein-ligand complexes will be invaluable to train the next-generation of deep learning models for docking.

To address this issue, we propose a novel method for the generation of synthetic protein-ligand complexes.
As opposed to existing methods, our approach contributes two major advancements.
First, we employ a range of diverse synthetic complex generation strategies including PDB fragmentation, de novo ligand design, and de novo protein design.
This ensemble improves the diversity of generated structures while also reducing bias that arises from the creation process.
Secondly, we implement a rigorous validation procedure that incorporates both physics-based and ML-based filtering.
This ensures that we retain only high-quality data that obeys known physics and are indistinguishable from experimental structures.

\section{Background and Related Work}
\subsection{Traditional Molecular Docking}
Docking is typically divided into two sub-tasks: sampling and scoring.
A search function is tasked with sampling low-energy ligand poses within the binding site while a scoring function is tasked with ranking these samples.
Existing search functions are largely based on well-studied sampling methods such as Monte Carlo and genetic algorithms paired with a scoring function \cite{altuntacs2016gpu}.
Scoring functions are generally categorized into physics-based, knowledge-based, and empirical methods, each relying on different theoretical and computational approaches to estimate binding affinity.
Physics-based methods estimate the energy of binding poses based on detailed physical molecular interactions.
Knowledge-based methods derive scoring functions from calculated statistics of known protein-ligand complexes, such as the potential mean force, while empirical methods use regression or other trained predictors to predict binding affinity from structural features of the docked pose \cite{fujimoto2022machine}.
Additionally, docking can be done with or without knowledge of a specific protein binding site, termed \textit{focused docking} and \textit{blind docking} respectively.
While most docking programs are intended for focused docking, they can be adapted for blind docking by pairing with a pocket prediction tool \cite{masters2023pocketnet}.

\subsection{DL-based Molecular Docking}
Initial applications of deep learning models to docking were aimed at improving the scoring function.
These methods used an existing docking program to generate the poses but improved the results by reranking them with a neural network scoring function.
Several different networks have been applied in this way including convolutional neural networks \cite{ragoza2017protein, mahmoud2020elucidating} and graph neural networks \cite{wang2021protein, townshend2021geometric}.
Additionally, many DL models have been developed specifically to improve binding affinity prediction accuracy compared to traditional scoring functions, and these enhanced predictive models can be effectively utilized for re-ranking docking poses \cite{ozturk2018deepdta,jiang2020drug,jones2021improved}.
While these approaches have proven to be powerful in improving the scoring of poses, they do not address the sampling problem.

Generative deep learning models have begun to replace the search function of classical docking models too.
There are several unique approaches including prediction of euclidean distance matrices \cite{mahmoud2020graph,masters2023deep,voitsitskyi2024artidock}, keypoint transformations \cite{ganea2021independent, stark2022equibind}, and trigonometry-aware neural networks \cite{lu2022tankbind}.
However, the most widely used approach has become diffusion-based models \cite{lu2024dynamicbind,nakata2023end,krishna2024generalized,yim2024diffusion} which began with the seminal work of DiffDock \cite{corso2022diffdock}.
DiffDock introduced a method for diffusing along the ligand conformational and orientational degrees of freedom, allowing for the flexible fitting of a ligand into its binding site.
The stochastic nature of diffusion models enables the generation of diverse binding poses.
While DiffDock was initially intended for blind docking to a rigid protein structure, it's architecture has been adapted in several works in order to perform focused docking \cite{guo2023diffdock,plainer2023diffdock} and flexible protein docking \cite{lu2024dynamicbind}.
One study investigated the performance of DL models for blind docking and found that they mostly excelled due to their pocket finding ability \cite{yu2023deep}.
Recently, the authors of DiffDock released an updated version featuring confidence bootstrapping, synthetic training data, and models scaled to tens of millions of parameters \cite{corso2024deep}.
The success of deep learning models like AlphaFold2 and RoseTTAFold at protein structure prediction led to the release of AlphaFold3, RoseTTAFold-AllAtom, and other so-called co-folding models which are capable of molecular docking small molecules as well as a host of other features \cite{abramson2024accurate,chai2024chai,wohlwend2024boltz,bytedance2025protenix,krishna2024generalized}.

\subsection{Synthetic Protein-Ligand Complexes}
\subsubsection{Ligand-Binding Protein Design}
Although the idea of generating synthetic protein-ligand complexes for training deep learning models is relatively new, it intersects with a number of earlier developments in de novo protein and ligand design.
For instance, the problem of ligand-binding protein design, where researchers are interested in creating a protein capable of binding a specific molecule with high affinity overlaps with our idea of generating synthetic high-quality complexes.
There are numerous applications for such a method, for instance in the development of antitoxins, sequestering agents, and antidotes for drug overdose \cite{tinberg2013computational,moretti2016rosetta}.
Initially, many attempts at ligand-binder design were unsuccessful, but advanced simulation and design tools like the Rosetta package, enabled some successes.
Other approaches rely heavily on existing PDB structures, by docking ligands onto proteins with high shape complementarity to the ligand \cite{polizzi2020defined, lu2024novo}.
Recently, some developments have been made to accelerate and enhance the success of these methods by using deep learning \cite{an2023novo, liu2024design}.

\subsubsection{De Novo Ligand Design}
Traditionally, virtual screening and experimental high-throughput screening were the predominant methods for identifying new lead compounds.
However, this is highly resource intensive as one must enumerate an enormous ligand library, most of which will fail to bind.
In de novo ligand design, a researcher is interested in designing a ligand from scratch, usually with some intended target and properties in mind.
This can be highly advantageous as it removes the costly screening procedure and can enable one to find high-quality, rationally-designed leads from the start.
While this topic has existed since the inception of rational design \cite{moon1991computer}, it has also seen numerous recent developments, particularly with the introduction of deep learning models for de novo ligand design \cite{krishnan2021novo,wang2022deep}.
In Pocket2Mol, researchers developed an E(3)-equivariant generative model capable of generating novel molecular structures given a 3D protein pocket \cite{peng2022pocket2mol}.
DiffSBDD takes a similar approach, enabling users to generate de novo ligand structure with a diffusion-based model conditioned on the pocket structure \cite{schneuing2024structure}.

\subsubsection{Generating Synthetic Training Sets}
The idea of generating synthetic data to assist in training deep learning models is not new, or unique to molecular modelling.
Synthetic data has been used to train models for image segmentation \cite{hinterstoisser2018pre,nowruzi2019much}, facial recognition \cite{hu2017frankenstein}, simulated environments \cite{zhang2019survey}, and more \cite{nikolenko2021synthetic,lu2023machine}.
In the context of machine learning for molecules, synthetic data has been utilized in several papers.
For example, in Voitsitskyi et. al. \cite{voitsitskyi2024augmenting}, they developed a novel data augmentation approach which involves placing favorably interacting residues around a ligand to mimic the distribution of interactions found within the PDB.
However, this approach showed no improvement in model accuracy, only in reducing the number of observed steric clashes within the predictions.
The approach was later modified and used to train ArtiDock, which showed an improvement compared to other physics-based and ML-based models for docking \cite{voitsitskyi2024artidock}.
In another work by Gao and Jia et. al. \cite{gao2023profsa}, they developed another approach for introducing synthetic data originating from the PDB.
They selected residues at random and removed them from the sequence, introducing a new binding pocket which is weakly associated with the short peptide segment that was removed.
Finally, in Corso and Deng et. al. \cite{corso2024deep}, they took inspiration from the van der mer technique introduced earlier in order to generate additional training data for their docking model.
Generally, it has been found that generating synthetic data through a variety of models or methods produces more robust downstream models \cite{van2023synthetic}.
This intuitively makes sense, as a single data generation approach may have certain limitations and biases, but when mixed alongside other data generation methods does not bias the downstream model as much.
We take inspiration from this finding to develop our synthetic data generation methodology.
By combining the results of several different data generation tools, we can avoid the bias present in any one dataset and build a robust dataset that covers a wide range of possible chemical space.



\section{Materials and Methods}

\begin{figure}[!ht]
    \centering
    \includegraphics[width=1.0\linewidth]{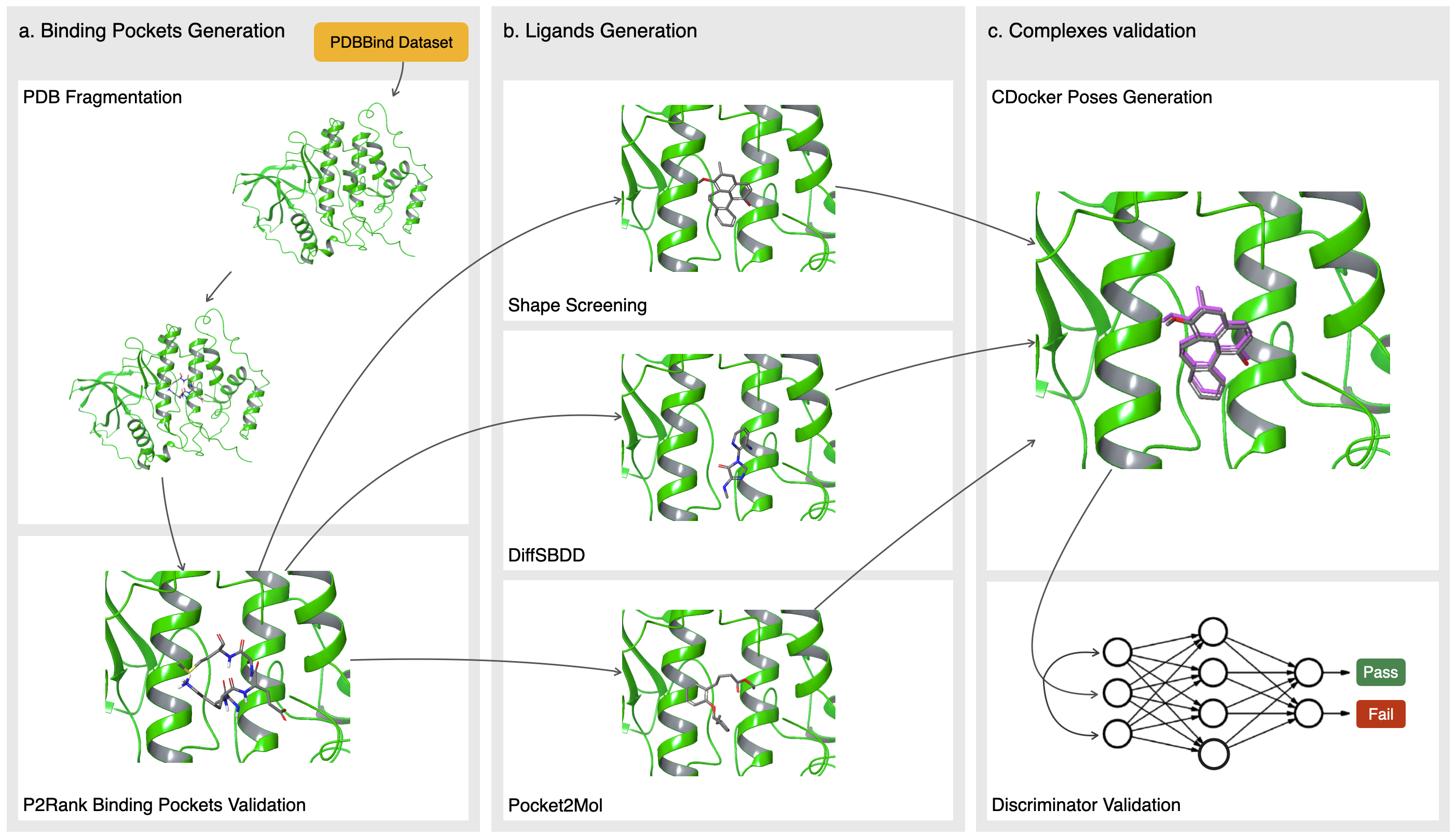}
    \caption{Overview of synthetic protein-ligand data generation workflow. a. Protein pocket definition and validation from experimental structures using fragmentation and P2Rank b. Ligand generation by shape screening, DiffSBDD, and Pocket2Mol methods c. Complex validation using CDocker redocking and neural network classification}
    \label{fig:workflow_overview}
\end{figure}

\subsection{Synthetic Data Ensemble Generation}
Rather than relying on a single method to generate our synthetic complexes, we decided to use an ensemble of diverse methods for synthetic data generation.
As each method has its own inherent limitations and is not capable of exploring the entirety of small molecule and sequence space, the ensembling of multiple methods allows us to generate even more diverse complexes, while avoid biases specific to one particular method.
For example, while one method may generate smaller fragments, another may generate larger ligands with many functional groups.
Combining both approaches gives the best of both worlds by allowing the model to work on molecules of various sizes and properties.

\subsubsection{PDB Fragmentation}
We begin by identifying protein structures containing between 100 and 1000 residues from the entire PDB database.
For each qualifying protein, our workflow selects multiple fragments of varying lengths, ranging from 1 to 10 residues.
This fragmentation involves randomly selecting starting positions within the protein sequence and extracting the specified number of residues to form each fragment.
Our workflow then extracts the selected fragment from the remainder of the protein structure.
This process creates an artificial gap in the sequence and forms a new pocket in the 3D structure.
The extracted polymer segment is considered the ligand, and the remainder of the protein, including the newly created pocket, is considered the receptor.
These fragments and their corresponding receptor structures are saved as separate PDB files in newly created directories.
The naming convention for these directories includes information about the original protein and the residue range of the fragment.
The putative pockets are then validated using the pocket detection tool P2Rank  \cite{krivak_p2rank_2018,jakubec_prankweb_2022,jendele_prankweb_2019,krivak_p2rank_2015,krivak_improving_2015}.
P2Rank works employing a random forest classifier to assess the ligandability of local chemical neighborhoods on the solvent-accessible surface of proteins.
Unlike traditional models, P2Rank does not rely on pre-existing pocket detection algorithms but directly predicts ligandable points and clusters them to identify potential binding sites, potentially discovering novel sites overlooked by geometric or template-based methods.
After obtaining the validated pockets, we employ three different methods to replace the removed peptide fragments with diverse, drug-like ligands.

\subsubsection{Shape Screening}
One of the main issues with the PDB fragmentation described above is that the chemistry found in protein fragments is limited and functional groups often found in drugs, such as halogens, sulfonamides, phosphates, heterocycles, charged atoms, etc. are missing.
Therefore, our first approach to replace the peptide with drug-like molecules involves using shape screening across a large library of small molecule ligands.
To this end, we perform shape-based screening using Schrödinger's Shape Screen GPU tool against a library of approximately 10 million diverse small molecules from the MolPort Screening Compound database \cite{sastry_rapid_2011}.
This step involves comparing the shape of our reference ligands (those isolated from the fragmentation step) against a large library of compounds to identify molecules with similar three-dimensional shape and pharmacophores.
After the shape screening, we run Smina \cite{koes_lessons_2013} on the top hits to obtain poses and minimize the top poses using OpenMM \cite{eastman_openmm_2023}.

\subsubsection{De-Novo Ligand Generation}

\paragraph{DiffSBDD}

Our second approach involves generating synthetic ligands for protein binding sites using the DiffSBDD tool introduced earlier \cite{schneuing_structure-based_2024}.
Within our workflow, we iterate through all previously validated protein fragments in our dataset and, in each case, analyze the spatial relationship between the protein and the existing ligand (our protein fragment), calculating distances between each atom in the protein and each atom in the ligand.
Protein residues within 8\AA\ of any ligand atom are considered part of the binding site definition provided to DiffSBDD.
DiffSBDD was then run using default settings on GPU to generate de novo ligands which occupy the synthetic binding pocket.
DiffSBDD is designed to produce high-affinity binders which forms multiple interactions with the protein in the binding site, although it is generally unknown or experimentally validated how reliable this ligand generation process is.
Therefore, we will later combine the synthetic datasets produced by the various methods and try to validate their fidelity through machine learning and physics-based methods.

\paragraph{Pocket2Mol}
Our workflow's third option takes a similar approach to generating synthetic molecules for protein binding sites using the Pocket2Mol tool \cite{peng_pocket2mol_2022}.
Within our workflow, we iterate through all previously validated protein fragments in our dataset and, in each case, calculate the center coordinates of the ligand (our protein fragment) by averaging the coordinates of all its atoms, which serves as a reference for Pocket2Mol's generation efforts.
Pocket2Mol is then run using default settings on GPU to generate the de novo ligand structures.

\subsection{Physics and ML-based Validation}
We used a combination of physics-based and ML-based validation in order to assess generated complexes and filter low-quality data.

\subsubsection{Redocking with CDocker}
In order to assess the physical validity of our generated data, we re-docked the ligands coming from each of the three generation methods using the force-field-based CDocker engine \cite{gagnon_flexible_2016}.
This redocking is a rigorous measure to ensure that the generated pose can be recapitulated based on physics alone.
If CDocker was able to reproduce a pose with RMSD <2\AA\ then this synthetic protein-ligand complex passes this phase of the quality control check.

\subsubsection{Neural Network Classifier}
To further our validation process beyond redocking, we developed a simple classifier model that is trained to predict whether a given complex was experimentally determined or generated via our procedure.
The idea is that if the discriminator, when well-trained, cannot differentiate a particular set of synthetic data from experimental data, it indicates these synthetically generated samples are of good quality and resemble experimentally determined structures.
To this end, custom fingerprints describing the protein pocket, ligand, and their interactions is calculated via Po-sco \cite{sellner2023quality} and is provided as input to the network.
A simple feed-forward network with three hidden layers of 256 weights, ReLU activation functions, and a sigmoid-activated output layer is trained to discriminate between the two classes (0 = experimental; 1 = synthetic).
The network was trained using the Adam optimizer with learning rate of 0.001 for a total of 100 epochs.
A balanced sampler was used to even sample positive and negatively labeled data, as to not bias the model towards one class or the other.
Following training, any synthetic data sample with a score above 0.5 was removed from the dataset, resulting in our final synthetic training sets.

\subsection{Retraining-based validation}

To further evaluate the quality of our synthetic data, we retrained two relatively straightforward models: Smina, an open-source molecular docking and scoring tool derived from AutoDock Vina, designed to offer improved support for minimization, scoring, and custom scoring functions in molecular docking tasks \cite{koes_lessons_2013}, and Gnina, an open-source molecular docking software that integrates a deep-learning-based scoring functions using convolutional neural networks to predict protein-ligand binding poses and affinities \cite{mcnutt_gnina_2021,ragoza_protein-ligand_2017}.
The two models were chosen to investigate the impact of synthetic data on both low-parameter and high-parameter models.
Smina only contains five parameters based on simplified physical interactions and can be fit with a few number of training samples.
Gnina is a deep neural network and contains thousands of trainable parameters that work on 3D density data of the binding site, allowing for much more finetuned scoring.
However, deep learning methods often take a plethora of data to train in order to gain strong generalization capabilities.

Initially, Smina is trained on a synthetic training set by fitting the weights of various physical terms.
The resulting custom Smina scoring function is then used to generate new poses for the synthetic complexes used in training, in order to prevent experimental bias from leaking into synthetic evaluation, and two test sets, specifically the PDBbind core set \cite{su_comparative_2019,li_assessing_2018,liu_forging_2017,liu_pdb-wide_2014,li_comparative_2014-1,li_comparative_2014,cheng_comparative_2009,wang_pdbbind_2005,wang_pdbbind_2004} and PoseBusters, a benchmark dataset designed to evaluate the physical validity of AI-based docking predictions through systematic quality checks that assess chemical consistency, stereochemistry, and geometric plausibility of protein-ligand complexes \cite{buttenschoen_posebusters_2024}. Unlike traditional docking benchmarks that focus primarily on pose accuracy, PoseBusters specifically identifies physically implausible conformations that may arise from deep learning-based methods.
For the generation of poses using Smina, we employed the following parameters: \verb|--exhaustiveness 50|, \verb|--num_modes 20|, and \verb|--seed 0|.
These generated poses from the training set are subsequently employed to retrain Gnina, thereby producing custom Gnina weights.
Finally, the trained Smina and Gnina models are evaluated on the two test sets.

\paragraph{Retraining Smina}
The Smina scoring function is a weighted linear combination of five terms based on simplified physical interactions: Gaussian attractive terms, repulsion, hydrophobic interactions, and non-directional hydrogen bonding.
A sixth term, the number of rotatable bonds, is used as an estimate for the impact of entropy on the affinity and therefore better ranking between different ligands.
However, since we are only interested in the pose ranking task here and not affinity prediction, the sixth term is not considered in our results.

The particle swarm optimization (PSO) algorithm \cite{kennedy_particle_1995,shi_modified_1998} was employed to determine the optimal weights that minimize a specified loss function.

In this instance, a success-rate-based training loss was employed, where the objective function maximizes the fraction of protein-ligand complexes for which the top-ranked pose (according to the reweighted scoring function) has an RMSD below 2.0 Å relative to the native binding pose. Specifically, the loss function is defined as:

\begin{equation}
\mathcal{L}(\mathbf{w}) = -\frac{1}{N}\sum_{i=1}^{N} \mathbb{I}(\text{RMSD}_i^* < 2.0 \text{ Å})
\end{equation}

where $N$ is the number of protein-ligand complexes, $\mathbf{w}$ represents the weights for the five Smina scoring terms (gauss1, gauss2, repulsion, hydrophobic, and non-directional hydrogen bonding), $\mathbb{I}(\cdot)$ is the indicator function, and $\text{RMSD}_i^*$ is the RMSD between the top-ranked pose for complex $i$ and its corresponding native pose. The top-ranked pose for each complex is determined by minimizing the weighted linear combination of Smina scoring terms: $\text{pose}_i^{\text{best}} = \arg\min_{\text{pose} \in \text{poses}_i} \sum_{j=1}^{5} w_j \cdot f_j(\text{pose})$, where $f_j(\text{pose})$ represents the $j$-th scoring term value for a given pose. The RMSD values used in this optimization are the original, untransformed geometric distances calculated directly from atomic coordinates without any scaling or normalization applied.

For the PSO algorithm, the upper and lower bounds for the weights were defined by assigning specific bounds to different interaction terms: the lower bounds were set to -1 and the upper bounds to 0 for the terms \verb|gauss1|, \verb|gauss2|, \verb|hydrophobic|, and \verb|non_dir_h_bond|, while the lower bound was set to 0 and the upper bound to 1 for the term \verb|repulsion|.

\paragraph{Retraining Gnina}

The preparation of poses for retraining Gnina involved converting files into the GNINATYPES format.
Following this, a comprehensive list of poses and corresponding proteins was compiled into a TYPES file, where the organization of all elements is controlled by a set of flags applicable during both training and inference.
In this context, the LABEL is used to classify each pose according to its RMSD value from the reference pose, following the classification scheme described in the Gnina paper.
According to this scheme, poses with an RMSD below 2\AA\ are classified as binders and assigned a label of 0, while those with RMSD values between 2\AA\ and 4\AA\ are considered ambiguous and remain unlabeled.
Poses exhibiting an RMSD greater than 4\AA\ are categorized as non-binders and given a label of 1 \cite{ragoza_protein-ligand_2017}.
The synthetic dataset was then divided into a training set and a validation set using a 70/30 random splitting strategy, with 70\% of the data allocated for training and the remaining 30\% reserved for validation.
The re-training of Gnina was conducted on a single GPU over 10,000 iterations utilizing a batch size of 128.
To ensure reproducibility, the seed was fixed at 123.
The training process employed the default2018 model, and the model's convergence was confirmed through  analyzing the loss progression across successive iterations.

\section{Results and Discussion}

\subsection{Complex Generation and Validation}

\begin{figure}[!h]
    \centering
    \includegraphics[width=1.0\linewidth]{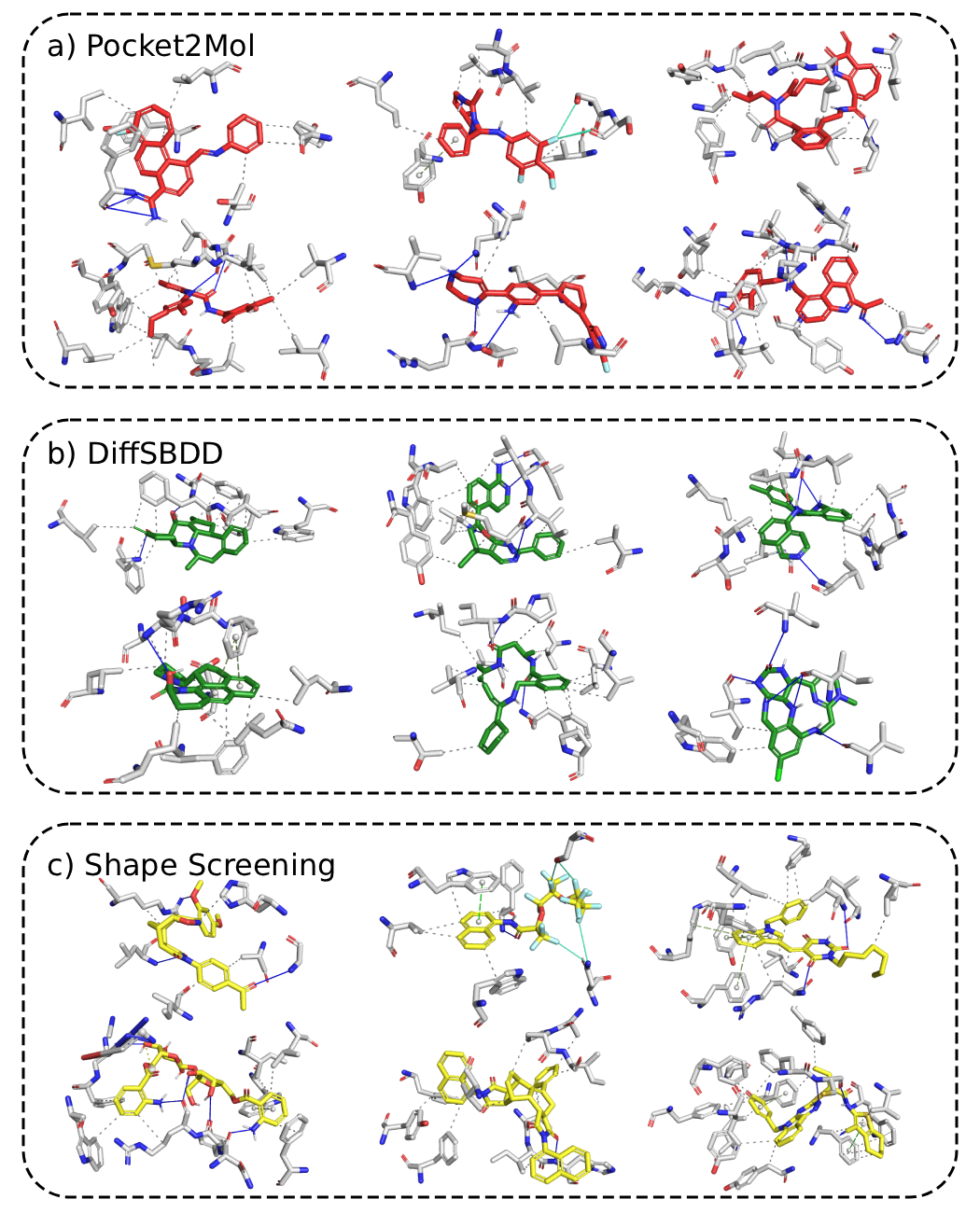}
    \caption{Selection of generated structures using each method. Protein-ligand interactions detected by PLIP \cite{} including hydrophobic (dashed gray), hydrogen (blue), halogen (cyan), and pi-stacking (gray sphere)}
    \label{fig:examples}
\end{figure}

\begin{figure}[!h]
    \centering
    \includegraphics[width=1.0\linewidth]{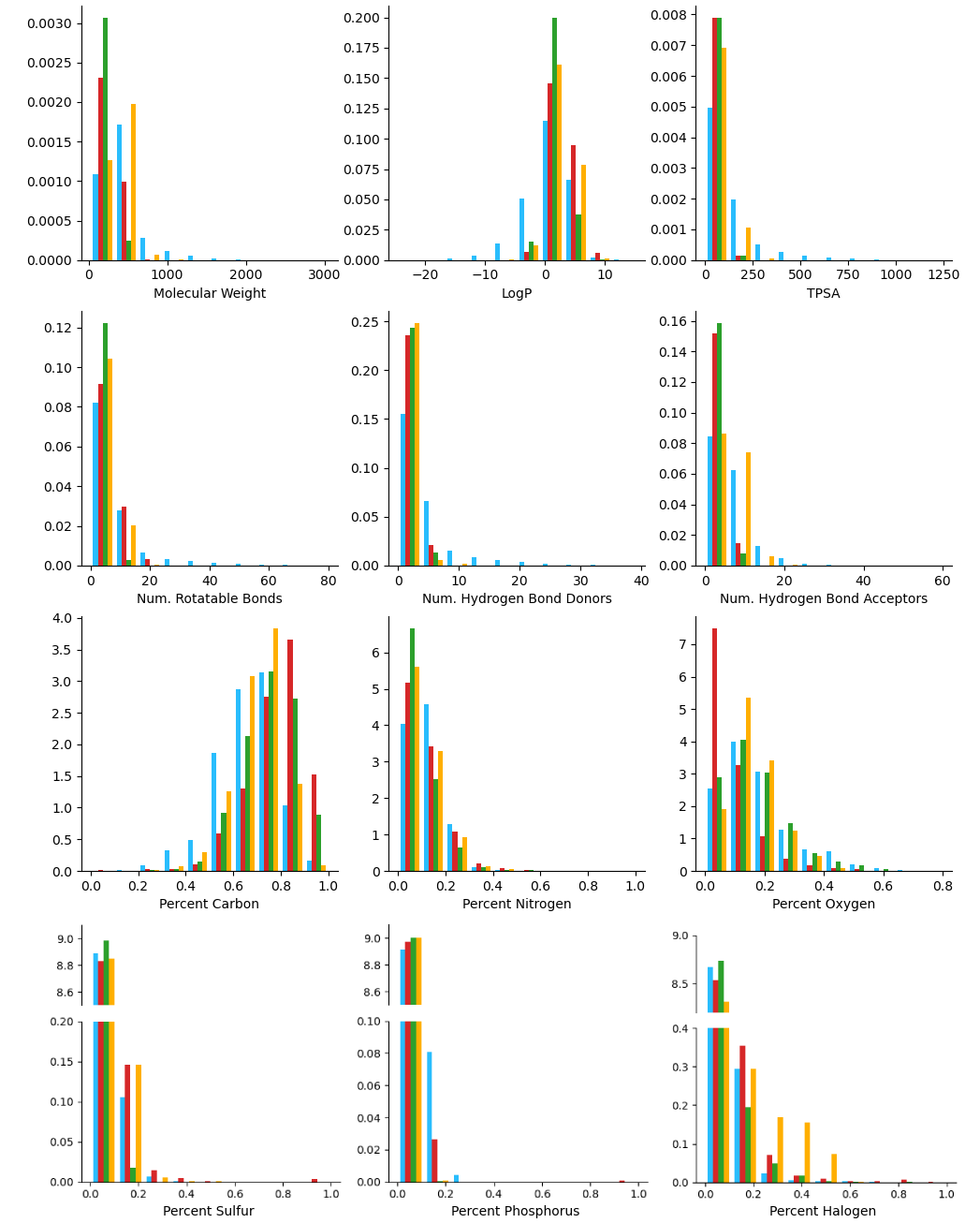}
    \caption{Distribution of ligand properties from each source: PDBBind (light blue), Pocket2Mol (red), DiffSBDD (green), shape screening (yellow). X-axis indicates the property value and y-axis indicates the density of that bin. Last row contains y-axis breaks since most ligands contain no sulfur, phosphorus, or halogen}
    \label{fig:properties}
\end{figure}

\begin{figure}[!h]
    \centering
    \includegraphics[width=0.85\linewidth]{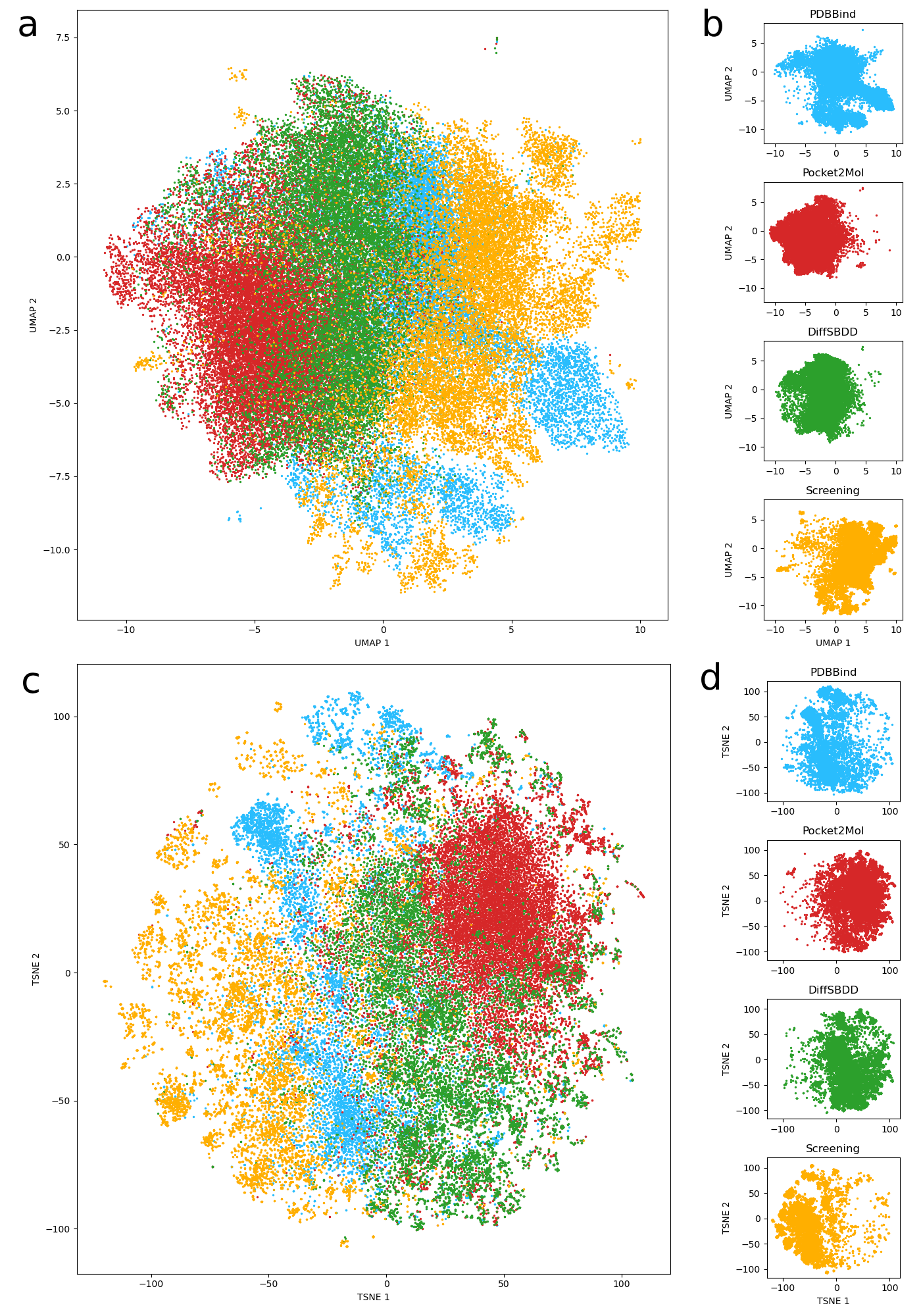}
    \caption{UMAP and TSNE projections of MoLFormer-XL embeddings for ligands from each source: PDBBind (light blue), Pocket2Mol (red), DiffSBDD (green), shape screening (yellow). A) UMAP projection of all sources. B) UMAP projections of each source individually. C) TSNE projection of all sources. D) TSNE projectsions of each source individually}
    \label{fig:umap_tsne}
\end{figure}

Visualized examples of synthetic protein-ligand complexes generated via each method can be seen in Figure \ref{fig:examples}.
All three methods are able to create ligands which satisfy shape complementarity and form favorable interactions with the synthetic binding pocket.
There is a good amount of diversity among each of the methods, producing ligands with a variety of sizes, chemical compositions, and functional groups.
Physical and chemical properties of generated ligands was analyzed in comparison to ligands occuring in the PDBBind Dataset (Figure \ref{fig:properties}).
In terms of molecular weight, the experimental ligands skew larger, with several very large peptidic ligands compared to our synthetic set.
The shape screening method also produced noticeably larger molecules than the two de novo methods due to the presence of these large ligands in the screening set.
In terms of LogP, our synthetic ligands are mostly positive and cover a different space than the PDBBind ligands which lean negative.
Total polar surface area (TPSA), number of rotatable bonds, hydrogen bond acceptors, and donors are all correlated and show the same trend observed with molecular weight.
The elemental composition of the ligands is similar among the different methods with some exceptions.
The PDBBind ligands are less likely to be rich in carbon (>80\%), while ligands generated using DiffSBDD and Pocket2Mol are more likely to contain that much carbon.
Pocket2Mol appears to have a preference against inserting oxygen atoms as more than half of the generated molecules do not contain any oxygen, while other methods often contain around 5-25\% oxygen.
Sulfur and phosphorus occurs rarely across all of the datasets, including PDBBind ligands.
However, all methods do generate at least one molecule containing these rarer organic elements.
Halogens also occur rarely, but each set of ligands contains at least 20\% halogenated compounds.
The shape screening set contains markedly more halogens due to the presence of polyfluorinated compounds in the screening set and the inability of the de novo methods to generate these types of ligands.
Finally, the chemical space of all four sets was visualized via dimensionality reduction of pre-trained molecular embeddings of the ligands, shown in Figure \ref{fig:umap_tsne}.
The two de novo methods, DiffSBDD and Pocket2Mol, largely occupy overlapping regions of chemical space.
This intuitively makes sense as both methods are using the same technology (generative neural networks) and both were optimized on similar PDB training data.
The shape screening ligands contain some overlap with the de novo ligands, but also occupy a different region of chemical space.
This can be explained by the previous property analysis which confirmed the presence of diverse ligands in the screening set, including large and exotic chemistry.
Ligands from PDBBind overlap with ligands from all other methods, but also contains ligands in an unexplored part of chemical space.
Many of these are the larger, peptidic ligands as none of our synthetic generation methods are capable of reliably producing these structures.

\subsubsection{CDocker Redocking-based System Selection}
\begin{figure}[!h]
    \centering
    \includegraphics[width=0.6\linewidth]{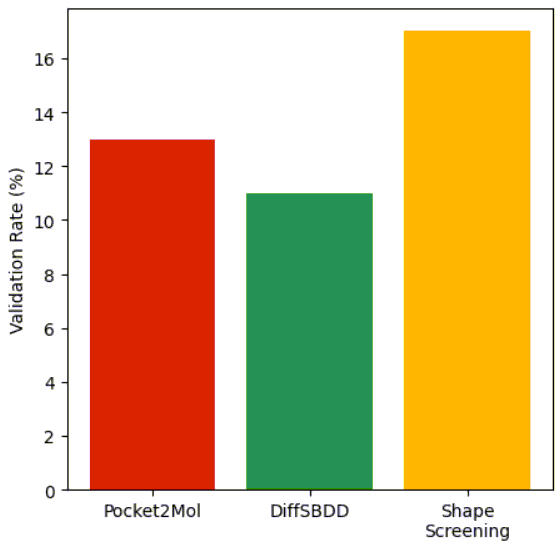}
    \caption{Results of CDocker redocking based validation}
    \label{fig:cdocker_results}
\end{figure}

Following the complex generation protocols described previously, two validation methods were applied to each of the generated systems: CDocker physics-based redocking and a discriminator neural network.
The results of the CDocker redocking on each of the generation methods is presented in Figure \ref{fig:cdocker_results}.
The results show that the generated complexes often fail the physics-based redocking procedure.
The passing rates were 13\%, 11\%, and 17\% for Pocket2Mol, DiffSBDD, and shape screening, respectively.
This was a foreseen outcome, as none of the generation methods involve rigorous physics-based protocols and therefore likely generates many false binders.
Nonetheless, these low quality complexes can be successfully filtered by this redocking protocol.
Generated structures which fail redocking (no CDocker pose within 2\AA\ RMSD) are removed from the set for further processing.

\subsubsection{Discriminator-based System Selection}
Following the physics-based validation via CDocker redocking, generated complexes were subjected to one more validation protocol: a classification neural network which attempts to filter real from synthetic complexes.
Initially, individual discriminator networks were trained on each set of generated complexes individually.
Then, all synthetic data was collected into a single dataset and used to train the discriminator used for filtering.
The results, presented in Table \ref{tab:discrim}, suggest that combining synthetic data from multiple sources produces the most robust dataset that is hardest to distinguish from real protein-ligand data.
This result agrees with our earlier assumption that combining synthetic data from multiple sources is advantageous since it removes the bias and distinguishing features of any single method.

\begin{table}[!htbp]
    \centering
    \begin{tabular}{cccc}
    \toprule
    \textbf{Generation Method} & \textbf{Loss} \\
    \midrule
    Shape Screening & 0.60 \\
    Pocket2Mol & 0.53 \\
    DiffSBDD & 0.58 \\
    Combined & 0.74 \\
    \bottomrule
    \end{tabular}
    \caption{Comparison of final loss of the neural network discriminator used for quality control. The value is an indirect indication of the quality of the synthetic data, as 0.0 indicates the network is able to completely detect real from fake and 1.0 indicates the network is unable to detect real from fake at all.}
    \label{tab:discrim}
\end{table}

\subsection{Retraining-based validation}

\subsubsection{Retraining Smina}

\begin{figure}[h]
    \centering
    \includegraphics[width=1\linewidth]{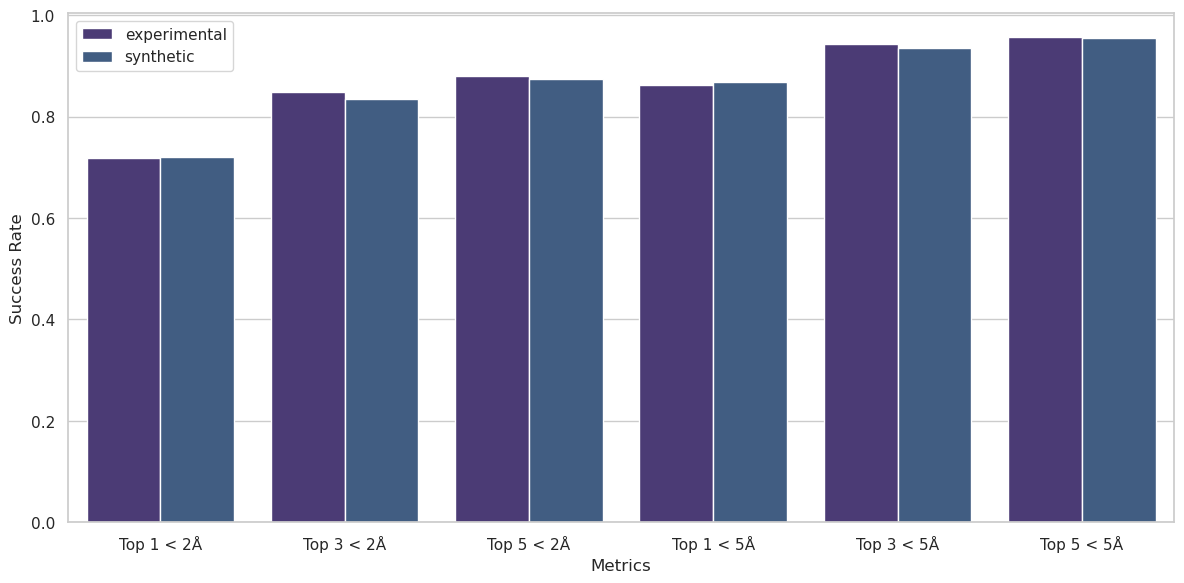}
    \caption{Comparison of the performance (success rate) of Smina models retrained on experimental data (violet) and synthetic data (blue) across multiple metrics.
    From left to right, bars indicate the proportion of top 1, top 3, and top 5 poses with an RMSD value less than 2\AA\, as well as the proportion of top 1, top 3, and top 5 poses with an RMSD value less than 5\AA}
    \label{fig:Smina_splits_performance_no_similarity_reduction}
\end{figure}

To evaluate the performance of the retrained Smina model, docking experiments were carried out on two test sets, namely the PDBBind core set and PoseBusters, with the success rate assessed across several metrics.
The results of this analysis, comparing both experimental and synthetic data, are depicted in Figure \ref{fig:Smina_splits_performance_no_similarity_reduction}.
Notably, the success rates observed are high and closely matched between the experimental and synthetic datasets across all evaluated metrics, demonstrating that retraining Smina solely on synthetic data yields performance on par with training on experimental data.
Table \ref{tab:smina_final_scoring_functions} details the weights employed in the original Smina scoring function as well as those used in the custom scoring functions derived from both experimental and synthetic datasets.
These weights are found to be highly similar regardless of the specific terms considered, highlighting the robustness of the results.
Furthermore, a duplicate training run was performed to assess the stability of the training process. The convergence to nearly identical parameters suggests that synthetic data captures similar underlying physics as experimental data, providing strong evidence for the physical realism of the generated complexes.

\begin{table}[!htbp]
    \centering
    \begin{tabular}{cccc}
    \toprule
    \textbf{Terms} & \textbf{original} & \textbf{synthetic} & \textbf{experimental} \\
    \midrule
    gauss(o=0,\_w=0.5,\_c=8) & -0.0356 & -0.037 & -0.056 \\
    gauss(o=3,\_w=2,\_c=8) & -0.0052 & -0.0054 & -0.0028 \\ 
    repulsion(o=0,\_c=8) & 0.8402 & 0.879 & 0.999 \\
    hydrophobic(g=0.5,\_b=1.5,\_c=8) & -0.0351 & -0.0366 & -0.0271 \\
    non\_dir\_h\_bond(g=-0.7,\_b=0,\_c=8) & -0.5874 & -0.6148 & -0.7169 \\
    \bottomrule
    \end{tabular}
    \caption{Comparison of the individual terms of the Smina scoring function between the original model, the model retrained on synthetic data, and the model retrained on experimental data.}
    \label{tab:smina_final_scoring_functions}
\end{table}

While the results are promising, the limitations of the current Smina model, particularly its simple scoring function with few physical terms, hinder our ability to achieve substantial improvements and effectively assess the usefulness of synthetic data.
Given these constraints, our subsequent step was to extend this approach to training Gnina, a more sophisticated model.

\newpage
\subsubsection{Retraining Gnina}

\begin{figure}[ht]
    \centering
    \includegraphics[width=1\linewidth]{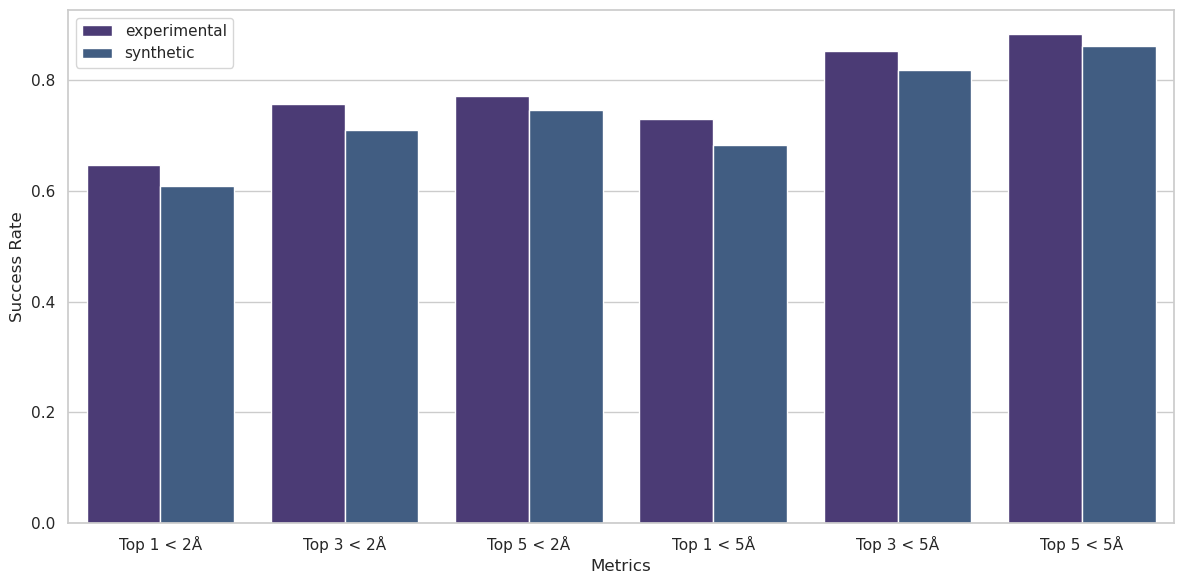}
    \caption{Comparison of the performance (success rate) of Gnina models retrained on experimental data (violet) and synthetic data (blue) across multiple metrics.
    From left to right, bars indicate the proportion of top 1, top 3, and top 5 poses with an RMSD value less than 2\AA\, as well as the proportion of top 1, top 3, and top 5 poses with an RMSD value less than 5\AA}
    \label{fig:Comparison_reranked_RMSD_no_similarity_reduction_Poses_4113}
\end{figure}

The results of retraining Gnina on both synthetic and experimental data are presented in Figure \ref{fig:Comparison_reranked_RMSD_no_similarity_reduction_Poses_4113}.
While the performance achieved using synthetic data is slightly lower, it remains closely aligned with the results obtained from experimental data across all evaluated metrics.

\begin{figure}[h]
    \centering
    \includegraphics[width=1\linewidth]{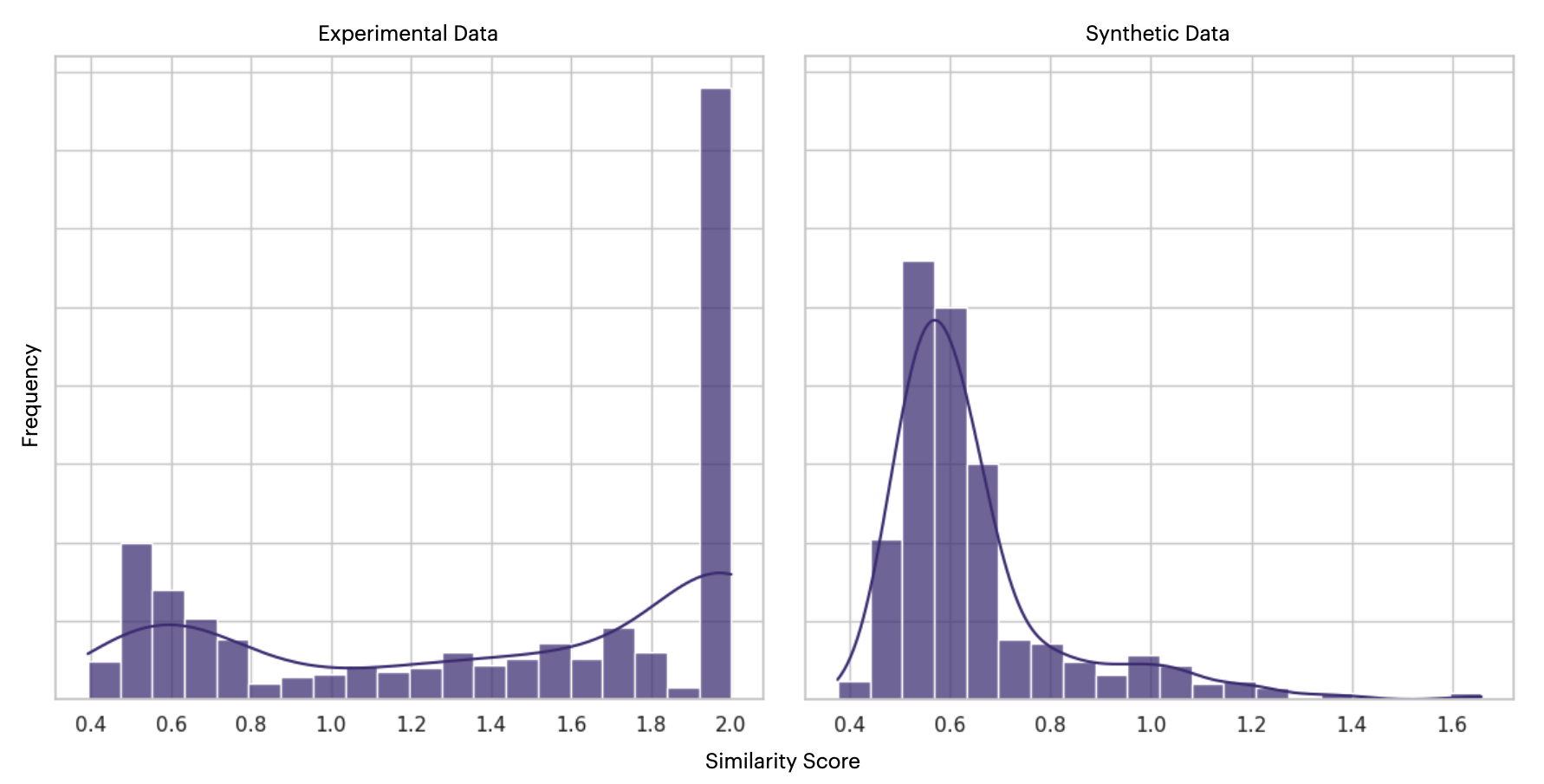}
    \caption{Binding site similarity analysis between test sets and training data using SiteMine's SP score. Distribution of maximum SP similarity scores (range 0-2, combining shape and pharmacophore similarity) for complexes from both test sets (PDBBind core set and PoseBusters) when compared against experimental training data (left) and synthetic training data (right). Each test set complex was compared against all complexes in the respective training set, and the maximum similarity score was recorded. The experimental training data shows substantially higher similarity to the test sets with many complexes achieving near-perfect matches (SP scores approaching 2.0), while synthetic training data demonstrates markedly lower similarity scores, providing quantitative support for the observed performance differences.}
    \label{fig:similarity_comparison}
\end{figure}

This marginal difference in performance may be explained by a higher degree of similarity between the complexes present in the experimental dataset and those in the test sets, compared to the similarity between the synthetic data complexes and the test sets. To quantify this hypothesis, we performed a systematic binding site similarity analysis using SiteMine's SP score \cite{Reim2024}, which combines shape similarity and pharmacophore similarity scores to assess pocket similarity on a scale from 0 to 2 (where individual shape and pharmacophore components each range from 0 to 1). For each complex in both test sets, we calculated the maximum SP score when compared against all complexes in the respective training sets. As shown in Figure \ref{fig:similarity_comparison}, the experimental training data exhibits substantially higher similarity to the test sets, with a significant fraction of binding pockets achieving near-perfect matches (SP scores approaching 2.0), while the synthetic training data demonstrates markedly lower similarity scores. This analysis provides quantitative support for the observed performance gap, as the experimental training data contains binding pockets that are more representative of those encountered in the evaluation benchmarks.

\subsection{Critical Discussion of Synthetic Data and Validation Strategies}

Despite the encouraging results obtained through retraining both Smina and Gnina on synthetic datasets, a number of caveats and limitations warrant careful consideration. First, the process of PDB fragmentation and shape‐based replacement can introduce bias into the synthetic ligand library toward chemotypes and physicochemical properties already present in the Protein Data Bank. Although the ensemble of generation strategies—including de novo diffusion‐based ligand design via DiffSBDD and Pocket2Mol—broadens chemical diversity, the reliance on protein fragments as structural anchors can skew ligand geometries toward small peptide‐like scaffolds. Consequently, downstream models may still underperform when confronted with small molecules possessing novel ring systems, charged moieties, or conformational flexibility that diverge substantially from the training fragments. This bias may modestly overestimate generalization performance when evaluating on benchmarks that inadvertently share pocket characteristics with the synthetic set.

The physics‐based and ML‐based validation steps mitigate but do not eliminate these biases. Redocking with CDocker and filtering by RMSD ensures that only ligands capable of refolding into low‐energy poses remain in the training corpus, yet this threshold does not guarantee accurate reproduction of subtle interaction networks such as water‐mediated hydrogen bonds or long‐range electrostatic complementarity. Likewise, the neural network discriminator offers an orthogonal filter to discard egregiously artificial complexes, but its resilience depends strongly on the balance between positive and negative examples and representativeness of the training examples. If the discriminator itself is trained on synthetic data that shares systematic artifacts—such as overly smooth pocket surfaces or idealized rotamer distributions—it may fail to detect less obvious but equally problematic generation errors. Thus, the retained high‐quality synthetic set might harbor minor structural inaccuracies that propagate into the learned docking functions of Smina and Gnina.

Moreover, the evaluation metrics applied in the retraining‐based validation emphasize static pose accuracy, predominantly through RMSD thresholds at 2Å and 4Å. While these metrics are standard, they do not fully capture the dynamic nature of protein–ligand binding. In particular, rigid‐body RMSD measures do not account for induced‐fit rearrangements of side chains or backbone movements that can be crucial for true binding affinity. The lack of explicit treatment of protein flexibility beyond minor side‐chain adjustments risks overfitting the models to rigid pockets and may limit performance on targets with shallow or allosteric sites. Similarly, success‐rate metrics alone do not reflect the energy landscape smoothness or the ranking robustness across multiple generated poses, which are key factors in virtual screening campaigns.

An additional factor contributing to the superior performance of experimental training data lies in the inherent structural adaptation present in crystallographic complexes. In experimental structures, the protein conformation has been optimized through co‐crystallization or structure refinement processes, resulting in side‐chain and backbone arrangements that are specifically adapted to accommodate the bound ligand. This induced‐fit adaptation includes subtle conformational adjustments, water molecule positioning, and local energy minimization that collectively optimize the protein–ligand interface. In contrast, synthetic complexes are typically generated using preformed protein structures without equivalent structural adaptation to the novel ligands. Consequently, the protein conformations in synthetic datasets may not exhibit the same degree of complementarity to their paired ligands, potentially leading to suboptimal interaction geometries and reduced binding affinity representations that could limit the effectiveness of models trained on such data.

Finally, retraining a physics‐based tool like Smina alongside a more expressive deep scoring model such as Gnina provides complementary validation but also highlights the known scope boundaries of each approach. Smina's relatively simple scoring function, tuned via particle swarm optimization, only mildly adjusts the original terms, suggesting that synthetic data alone may be insufficient to discover fundamentally new physics contributions. Conversely, Gnina's slight performance gap when trained solely on synthetic data underscores that structural novelty and realistic interaction patterns remain scarcer in generated sets than in experimental complexes. This gap may be partly attributed to our synthetic methods’ inability to reliably produce the larger peptidic ligands present in experimental datasets. Additionally, further improvements in generative methodologies, perhaps through integrating explicit solvent modeling or adversarial training to penalize unphysical contacts, are necessary to fully match the predictive power conferred by real protein–ligand structures.

In summary, while our synthetic data ensemble and the accompanying validation pipeline represent a substantial advance in expanding the training corpus for docking models, careful attention must be paid to residual biases, the sufficiency of structural realism, and the adequacy of evaluation metrics. Addressing these challenges through enhanced generative diversity, more rigorous physics‐informed screening, and dynamic validation assays will be critical to ensure that synthetic datasets can robustly substitute for, rather than merely supplement, scarce experimental complexes.




\section{Conclusion}

In this work, we address a critical bottleneck in the application of deep learning to molecular docking: the lack of sufficient high‐quality protein–ligand complex data.
By introducing a novel ensemble-based approach to generate synthetic protein-ligand complexes, we demonstrate that it is possible to procedurally expand the available training datasets while retaining a high degree of structural and physical plausibility.
Our methodology combines diverse generation strategies including PDB fragmentation, shape screening, and de novo ligand design via diffusion models with a rigorous validation framework encompassing physics‐based redocking, a neural network discriminator, and retraining‐based performance assessments.

Retraining both a physics‐based tool (Smina) and a deep learning‐enhanced model (Gnina) on these synthetic sets yields docking success rates closely aligned with those obtained using experimental complexes, underscoring the promise of synthetic data to enhance deep molecular docking pipelines.
Nevertheless, our critical evaluation highlights remaining limitations in generative diversity, potential biases toward peptide‐derived ligands, and the challenges of fully capturing protein flexibility and dynamic binding phenomena.

Future work will focus on integrating more advanced generative architectures and on developing benchmarks that probe applications to allosteric and highly flexible targets.
By systematically refining both data generation and validation procedures, we aim to pave the way for next‐generation docking models that are not constrained by experimental data scarcity and that can generalize across the proteome with high fidelity.

\section{Statements and Declarations}
Competing Interests: The authors declare no conflict of interest.


\bibliography{manuscript}
\bibliographystyle{unsrt}

\end{document}